\documentclass[aps,prl,reprint,preprintnumbers,superscriptaddress,amsmath,amssymb,bibnotes,longbibliography]{revtex4-2}

\usepackage{graphicx}
\usepackage{bm}
\usepackage{physics} 
\usepackage[colorlinks,linkcolor=blue,anchorcolor=blue,citecolor=blue,urlcolor=blue,filecolor=blue,menucolor=blue,runcolor=blue]{hyperref}

\graphicspath{{figures/}}

\begin{document}
	
	
	\title{Magnetic and electric properties of the metallic kagome antiferromagnet CrRhAs}
	
	\author{Franziska Breitner}
	\affiliation  {Experimental Physics VI, Center for Electronic Correlations and Magnetism, University of Augsburg, 86159 Augsburg, Germany}
	
	\author{Bin Shen}
	\email{bin.shen@physik.uni-augsburg.de}
	\altaffiliation{Present address: National Laboratory of Solid State Microstructures, School of Physics and Collaborative Innovation Center of Advanced Microstructures, Nanjing University, Nanjing, China}
	\affiliation  {Experimental Physics VI, Center for Electronic Correlations and Magnetism, University of Augsburg, 86159 Augsburg, Germany}

	\author{Anton Jesche}
	\affiliation  {Experimental Physics VI, Center for Electronic Correlations and Magnetism, University of Augsburg, 86159 Augsburg, Germany}
	
	\author{Alexander A. Tsirlin}
	\email{altsirlin@gmail.com}
	\affiliation{Felix Bloch Institute for Solid-State Physics, University of Leipzig, 04103 Leipzig, Germany}
	
	\author{Philipp Gegenwart}
	\email{philipp.gegenwart@physik.uni-augsburg.de}
	\affiliation{Experimental Physics VI, Center for Electronic Correlations and Magnetism, University of Augsburg, 86159 Augsburg, Germany}
	
	\date{\today}

	\begin{abstract}
		CrRhAs is an antiferromagnetic kagome metal predicted to host a nontrivial spin texture with vector spin chirality [Huang \textit{et al.}, \textit{npj Quantum Mater.} \textbf{8}, 32 (2023)]. We report the synthesis and basic characterization of CrRhAs single crystals, which exhibit an antiferromagnetic transition with $T_{\rm N}$ = 150~K, evidenced by electrical transport, heat capacity, and magnetization measurements. Hall resistivity varies linearly with magnetic field, i.e., there is no nonlinear Hall contribution. Intriguingly, the Hall coefficient changes sign between the configurations of $j \parallel ab, H \perp ab$ and $j \parallel c, H \perp c$, which is likely connected to a peculiar topology of the Fermi surface. Furthermore, for $j \parallel ab$, the Hall coefficient shows a pronounced and continuous enhancement below $T_{\rm N}$, signaling a significant reconstruction of the Fermi surface or an extra scattering from the magnons. Our results offer guidance for exploring anomalous electric transport phenomena in exotic magnetic systems.
	\end{abstract}
	
	\maketitle
	
	
	\section{I. Introduction}
	
	Kagome metals have emerged as a pivotal platform for exploring frustrated geometry and its interplay with electronic band topology. The two-dimensional kagome lattice can give rise to flat bands, van Hove singularities, and topological band crossings \cite{13WanPRB, 13KiePRL, 24ProPRB} that have been indeed observed in nonmagnetic kagome metals as well as magnetic kagome metals with ferromagnetic in-plane order~\cite{Wang2023,Sante2026}. Antiferromagnetic kagome lattice potentially leads to further interesting electronic properties because of various noncollinear spin configurations expected in this setting~\cite{24WanAMC}, but only a few kagome metals with antiferromagnetic ground states have been reported to date.
	
	
	In real kagome materials, additional factors can influence their electronic properties and magnetism. For example, interlayer couplings between the kagome planes and distortions of the ideal kagome lattice are commonly present in many of the kagome compounds. The ZrNiAl-type structure (space group $P\bar{6}2m$) represents a prominent family of intermetallic compounds featuring a distorted kagome lattice, where the transition-metal or rare-earth atoms occupy positions that deviate from the perfect symmetry of the ideal kagome network. Unlike the ideal kagome lattice, the twisted kagome geometry of the ZrNiAl structure incorporates two types of triangles rotated oppositely around the crystallographic $c$-axis, leading to a loss of inversion symmetry. This structural distortion profoundly modifies the electronic band structure and magnetism and may lead to new emergent quantum phenomena, such as novel spin textures and anomalous Hall effect (AHE) observed in HoAgGe~\cite{20ZhaoS, 24ZhaNP}.      
	
	CrRhAs crystallizes in the ZrNiAl-type structure, where the magnetic Cr atoms form a twisted kagome lattice in the $ab$-plane~\cite{deyris1974}. Early studies on polycrystalline samples of this material revealed an antiferromagnetic transition at 
	$T_{\mathrm{N}} \approx 165$~K~\cite{OHTA1990171,95SatJMMM}, accompanied by a minute 
	expansion of the $c$-axis and an equally small contraction of the $ab$-plane at $T_{\mathrm{N}}$~\cite{95SatJMMM}. Magnetization measurements showed a linear field dependence up to 15~T in the antiferromagnetic phase~\cite{92KanPB}. Hydrostatic pressure experiments demonstrated a slight suppression of $T_{\mathrm{N}}$ with increasing pressure~\cite{95SatJMMM}. Recent density functional theory (DFT) calculations confirm antiferromagnetic exchange couplings and predict spin textures where individual triangles adopt $120^{\circ}$ order either with positive or with negative vector chirality, and suggest the possible existence of nontrivial topological features in the electronic band structure of CrRhAs~\cite{23HuaQM}.
	
	Here, we report the successful growth of CrRhAs single crystals and present their fundamental properties alongside Hall effect measurements. Our study identifies CrRhAs as a correlated kagome antiferromagnet, likely adopting a coplanar noncollinear magnetic structure. No nonlinear Hall effect is observed in either the paramagnetic or antiferromagnetic state. Instead, the sign of the Hall coefficient reverses from positive for in-plane ($ab$-plane) current to negative for out-of-plane ($c$-axis) current configurations. Moreover, the Hall coefficient undergoes a dramatic change below the N\'eel temperature. This behavior suggests an anisotropic reconstruction of the Fermi surface or scattering rate driven by antiferromagnetic ordering.
	
	\section{II. Methods}
	
	Single crystals of CrRhAs were synthesized using a bismuth flux method. The starting materials were weighed in a stoichiometric ratio (Cr:Rh:As:Bi = 1:2:1:15) and placed inside a standard Canfield crucible set consisting of two Al$_2$O$_3$ crucibles and a strainer. This set was sealed inside a quartz tube under 200~mbar argon pressure. The crystal growth was performed inside a standard Muffle furnace placed under the fume hood. The sample was heated to 600~$^\circ$C over the course of 6~h and cured at this temperature for another 6~h. The temperature was then increased to 1100~$^\circ$C over 3~h and held for 2~h, followed by a slow cooldown to 650~$^\circ$C with a cooling rate of 2.5~$^\circ$C/h.  The crystals were then separated from the flux by centrifugation. Any remaining flux was removed by hydrochloric acid etching. The chemical composition and phase purity were verified by energy-dispersive x-ray spectroscopy (EDX) and x-ray diffraction (XRD). The XRD data were collected using CuK$_{\alpha}$ radiation on several needle-like crystals placed on a flat sample holder. 
	
	Additionally, high-resolution powder XRD data~\cite{xrd-data} were collected at the ID22 beamline of ESRF (Grenoble, France) using the wavelength of 0.4\,\r A and multi-analyzer detector setup~\cite{fitch2023}. A portion of crystals together with a small amount of residual flux was ground into fine powder and loaded into thin borosilicate glass capillaries that were spun during the measurement. Sample was cooled with the He-flow cryostat. \texttt{Jana2006} was used for the structure refinement~\cite{jana2006}.
	
	Electrical resistivity was measured in a Physical Property Measurement System (PPMS, Quantum Design) using a standard four-probe configuration. Heat capacity was determined via the relaxation method in PPMS. Magnetization measurements were performed in an MPMS3 SQUID magnetometer (Quantum Design). 
	
	For Hall effect measurement, three samples (S3, S4, and S5) were prepared with orthogonal current-field geometries. Samples S3 and S4 were measured with the current applied along the crystallographic $c$-axis, whereas S5 was measured with the current within the $ab$-plane. To optimize sensitivity, all samples were polished to thicknesses of 5--20~$\mu$m prior to measurement.
	
	Density-functional theory (DFT) band-structure calculations were performed in the \texttt{FPLO}~\cite{fplo} and \texttt{VASP}~\cite{vasp1,vasp2} codes using the experimental crystal structure at 50\,K. The Perdew-Burke-Ernzerhof version of the exchange-correlation potential~\cite{pbe96} was applied. 
	
	\section{III. Experimental results}
	
	\begin{figure}
		\includegraphics[angle=0,width=0.48\textwidth]{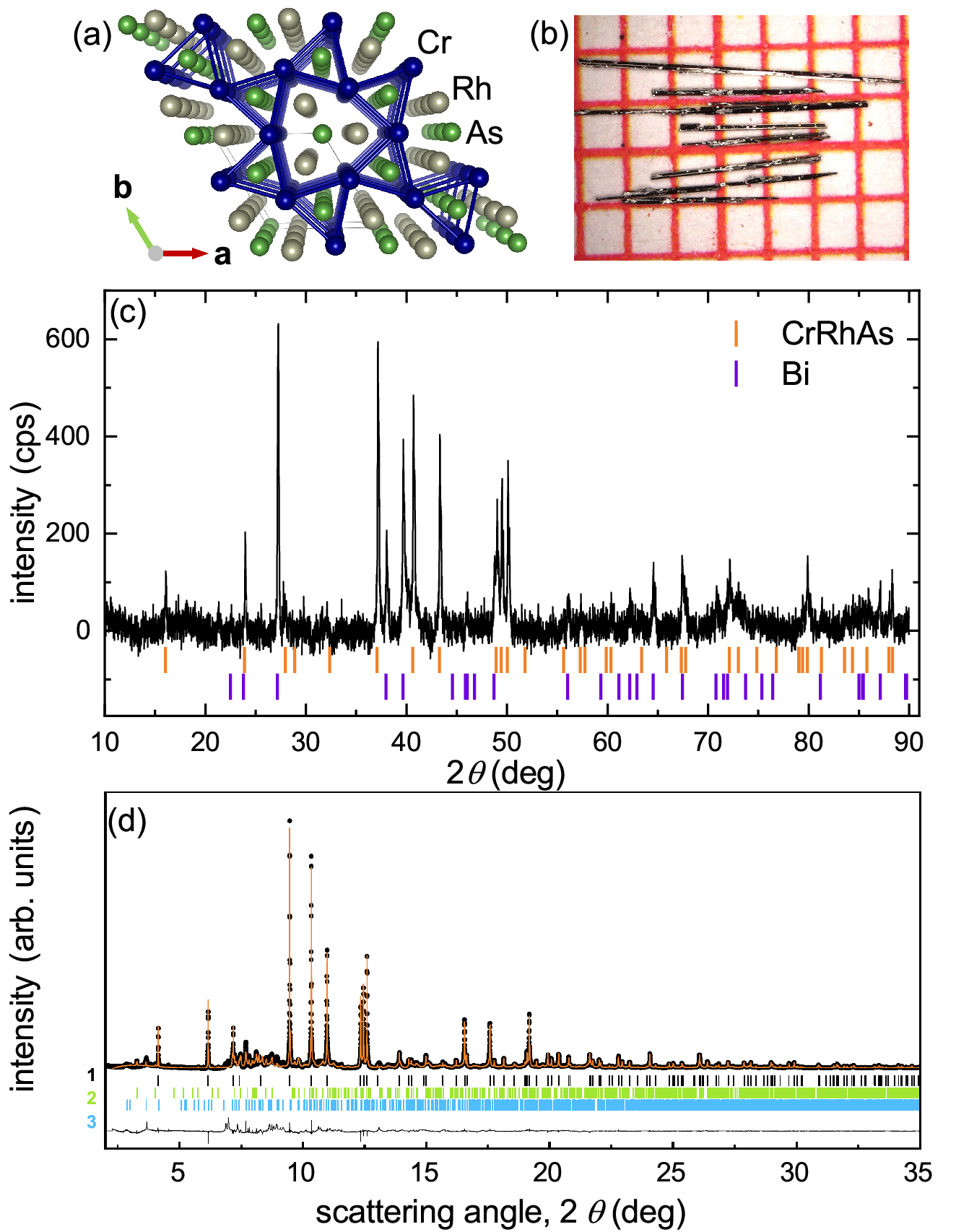}
		\vspace{-12pt} \caption{\label{XRD} (a) Crystal structure of CrRhAs. (b) Image of single crystals of CrRhAs. One grid square represents 1~mm. (c) Powder XRD pattern of CrRhAs collected on several needle-like single crystals placed on a flat holder. (d) Rietveld refinement for the high-resolution XRD data collected at 250\,K. Tick marks show peak positions for CrRhAs (1) and the impurity phases of RhBi$_3$ (2) and Rh$_3$Bi$_{14}$ (3).}
		\vspace{-12pt}
	\end{figure}
	
	\begin{figure}[!b]
		\includegraphics{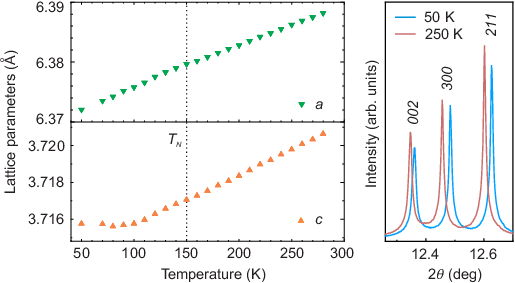}
		\caption{\label{fig:xrd}
			Left: temperature dependence of the lattice parameters of CrRhAs. Right: a snapshot of the high-resolution XRD patterns at 50 and 250\,K shows persistent hexagonal symmetry below $T_{\rm{N}}$.
		}
	\end{figure}
	
\begin{figure}
		\includegraphics[angle=0,width=0.48\textwidth]{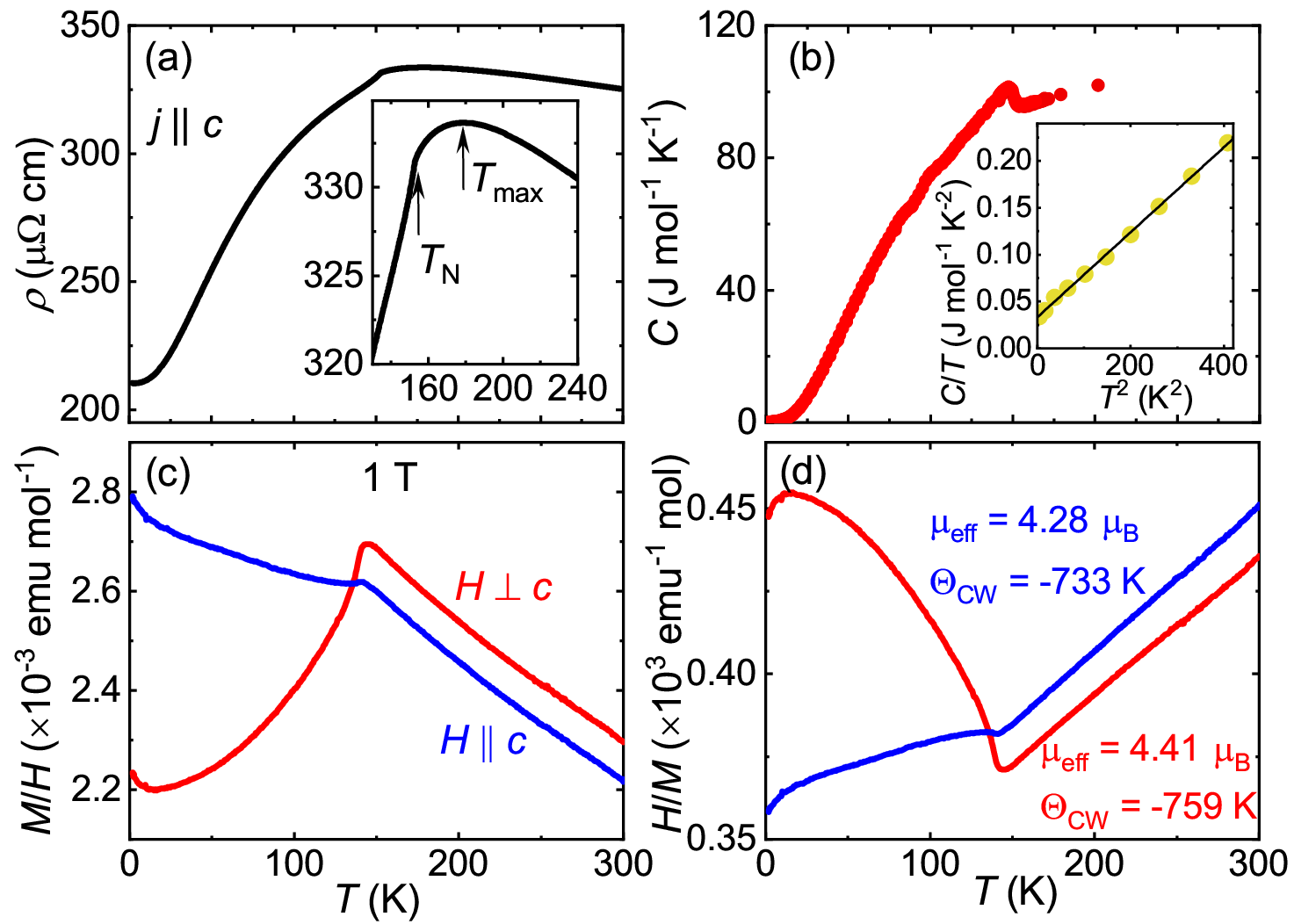}
		\vspace{-12pt} \caption{\label{CRAPro} (a) Temperature dependence of CrRhAs resistivity, with current applied along the $c$-axis. Arrows in the inset mark the antiferromagnetic transition temperature $T_{\rm{N}}$ and the temperature of the resistivity maximum $T_{\rm{max}}$, respectively. (b) Temperature-dependent heat capacity $C(T)$ of CrRhAs. The inset shows the fit to the low-temperature data with the formula $C/T = \gamma + \beta T^2$. Temperature dependence of (c) magnetic susceptibility and (d) inverse susceptibility with magnetic field applied along the $ab$-plane and $c$-axis, respectively. The Curie-Weiss fit to the high-temperature part of the data yields the effective moment of 4.28 (4.41)~$\mu_{\rm B}$ and the Curie-Weiss temperature of $-733$\,K ($-759$\,K) for the field applied along the $c$-axis ($ab$-plane).}
		\vspace{-12pt}
	\end{figure}

	Figure~\ref{XRD}(a) displays the crystal structure of CrRhAs, featuring chains of atoms aligned with the crystallographic $c$-axis. The single crystals exhibit a needle-like morphology oriented along the $c$-axis, with typical lengths reaching several millimeters [Fig.~\ref{XRD} (b)]. EDX measurements return the Cr:Rh:As ratio of 0.98(2):0.98(2):1.04(2) in a good agreement with the anticipated CrRhAs composition. Powder XRD further confirms the target hexagonal crystal structure. Trace amounts of elemental Bi were also detected in the XRD pattern when XRD was performed on multiple crystals extracted from the crucible without removing the residual Bi flux after centrifugation. This bismuth impurity is confined to the crystal surface, as confirmed by the absence of superconducting transitions attributable to bismuth in the resistivity measurements [Fig.~\ref{CRAPro} (b)] of carefully surface-cleaned CrRhAs single crystals.    
	
	Additionally, we performed high-resolution synchrotron XRD on a portion of crystals ground into fine powder and loaded into a glass capillary. As a larger amount of sample was needed for filling the capillary, the crystals were not pre-selected, and the sample contained traces of RhBi$_3$ and Rh$_3$Bi$_{14}$ phases from the flux [Fig.~\ref{XRD} (d)].
	High-resolution XRD data collected between 50 and 280\,K show a regular thermal expansion with no detectable elastic anomaly at the magnetic transition (Fig.~\ref{fig:xrd}). Not only the absence of the anomaly, but also the magnitude of thermal expansion are different from those reported in Ref.~\cite{95SatJMMM}. It may be a result of a different stoichiometry of the sample used in that study. Atomic coordinates are almost unchanged with temperature, whereas atomic displacement parameters gradually decrease on cooling (Table~\ref{tab:structure}). 
	
	The refinement with full site occupancies converged with the residuals of $R_I=0.022$ and $R_p=0.089$. Introducing mixed Cr/Rh site occupancies with the fixed Cr:Rh ratio of 1:1 according to EDX yields about 7\% site mixing and brings the residuals down to $R_I=0.021$, $R_p=0.087$. Although small amounts of site mixing can not be excluded, this effect is clearly less pronounced compared to CrPdAs where 28\% site mixing was reported~\cite{roy1979}. None of the XRD peaks split upon cooling (Fig.~\ref{fig:xrd}), suggesting that the $P\bar 62m$ symmetry is preserved in the magnetically ordered state of CrRhAs.   
	
	\begin{table}
		\caption{\label{tab:structure}
			Atomic coordinates and atomic displacement parameters $U_{\rm iso}$ (in $10^{-2}$\,\r A$^2$) for CrRhAs at 50\,K (upper line) and 250\,K (lower line). The lattice parameters are $a=6.37204(4)$\,\r A and $c=3.71576(3)$\,\r A at 50\,K vs. $a=6.38623(4)$\,\r A and $c=3.71979(3)$\,\r A at 250\,K. The refined Cr occupancy at the Rh site (and, equivalently, Rh occupancy at the Cr site) is 7(1)\%.}
		\begin{ruledtabular}
			\begin{tabular}{cccccc}
				&      & $x$       & $y$ & $z$       & $U_{\rm iso}$ \smallskip\\
				Cr  & $3g$ & 0.5947(7) & 0   & $\frac12$ & 0.8(1)        \\\smallskip
				&      & 0.5939(7) & 0   & $\frac12$ & 1.0(1)        \\
				Rh  & $3f$ & 0.2580(3) & 0   & 0         & 0.35(5)       \\\smallskip
				&      & 0.2578(3) & 0   & 0         & 0.80(5)       \\
				As1 & $1b$ & 0         & 0   & $\frac12$ & 0.7(1)        \\\smallskip
				&      &           &     &           & 1.1(1)        \\
				As2 & $2c$ & $\frac13$ & $\frac23$ & 0   & 0.7(1)        \\
				&      &           &           &     & 1.1(1)        \\
			\end{tabular}
		\end{ruledtabular}
	\end{table}
	
	Figure~\ref{CRAPro}(a) shows the temperature-dependent electrical resistivity $\rho(T)$ of CrRhAs with current applied along the $c$-axis. Upon cooling from room temperature, the resistivity initially increases slightly, reaching a broad maximum at approximately 180~K. Below this temperature, a distinct kink appears at 150~K, signaling the onset of antiferromagnetic ordering. The antiferromagnetic phase transition is further evidenced by a pronounced $\lambda$-like anomaly in the heat capacity, as shown in Fig.~\ref{CRAPro}(b). At low temperatures, the heat capacity follows the conventional expression $C(T) = \gamma T + \beta T^3$, where $\gamma T$ represents the electronic contribution and $\beta T^3$ describes the phonon contribution. Our analysis yields a substantial Sommerfeld coefficient of $\gamma = 33$\,mJ\,mol$^{-1}$\,K$^{-2}$, along with $\beta = 0.45$\,mJ\,mol$^{-1}$\,K$^{-4}$. From the $\beta$ coefficient, we calculate the Debye temperature of $\Theta_{\rm{D}} = 232$~K for CrRhAs.  
	
	\begin{figure}
		\includegraphics[angle=0,width=0.35\textwidth]{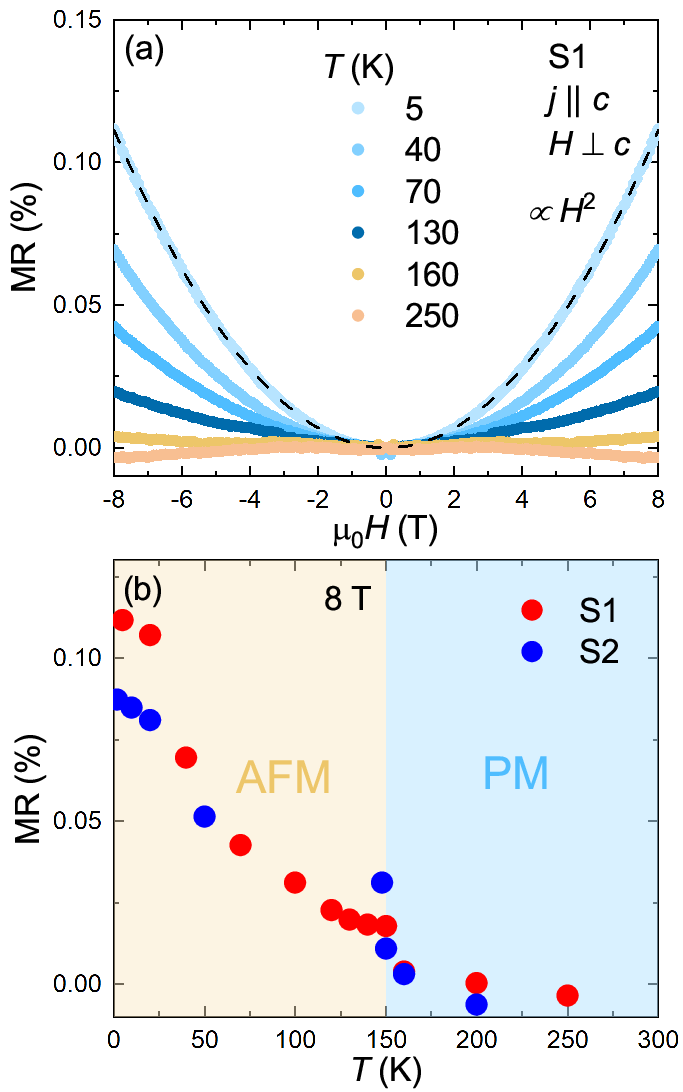}
		\vspace{-12pt} \caption{\label{MR} (a) Magnetoresistance $[\rho (H)-\rho (0)]/\rho (0)$ of CrRhAs at various temperatures of sample S1. In the antiferromagnetic state (below 150~K), MR varies as $H^2$, indicated by the dashed line. (b) Temperature dependence of the magnetoresistance at 8~T, $[\rho ({\rm 8~T})-\rho (0)]/\rho (0)$, of sample S1 and S2.}
		\vspace{-12pt}
	\end{figure} 
	
	Figure~\ref{CRAPro}(c) shows the temperature-dependent magnetic susceptibility of CrRhAs, with field applied along different directions. Above the magnetic transition at 150~K, the dc magnetic susceptibility follows the Curie-Weiss behavior [Fig.~\ref{CRAPro}(d)], yielding the effective moment $\mu_{\rm{eff}}$ = 4.28 (4.41)~$\mu_{\rm{B}}$ and the Curie-Weiss temperature $\theta_{\rm{CW}}$ = $-733$\,K ($-759$\,K) for $H\!\!\parallel\!\! c$ ($H\!\!\parallel\!\! ab$), suggesting the presence of localized Cr$^{3+}$ magnetic moments ($\mu_{\rm eff}=3.87$\,$\mu_{\rm{B}}$ assuming the spin-only magnetic moment) and only a weak anisotropy of their antiferromagnetic interactions. Below $T_{\rm{N}}$, magnetic susceptibility exhibits an anisotropic behavior typical of an antiferromagnet. For $H \perp c$, it shows a broad maximum at 150~K, decreasing to $\sim$80\% of this value at 2~K. In contrast, for $H \parallel c$, we observe a modest suppression near $T_{\rm{N}}$ followed by an upturn at lower temperatures. This characteristic temperature dependence is reminiscent of noncollinear antiferromagnets \cite{12JohPRL,19JinPRB,19HudJPCM}. It suggests that the spins lie in the $ab$ plane but do not align with one specific in-plane direction.
	
	Figure~\ref{MR}(a) presents the magnetoresistance (MR) of CrRhAs, defined as MR = $[\rho(H) - \rho(0)]/\rho(0)$, measu with current along the $c$-axis. In the paramagnetic state, the system exhibits weak negative MR. Below the antiferromagnetic transition, the MR becomes positive and follows a quadratic field dependence, which can be explained by the localized-spin scattering to the conduction electrons through a weak $s-d$ interaction \cite{73YamJPSJ}. The temperature evolution of MR of two samples at 8~T is shown in Fig.~\ref{MR}(b). While the paramagnetic state displays only small, weakly temperature-dependent MR values, a marked enhancement occurs upon entering the antiferromagnetic phase in both samples. Below $T_{\rm{N}}$, the MR increases with decreasing temperature.
	
	\begin{figure}
		\includegraphics[angle=0,width=0.48\textwidth]{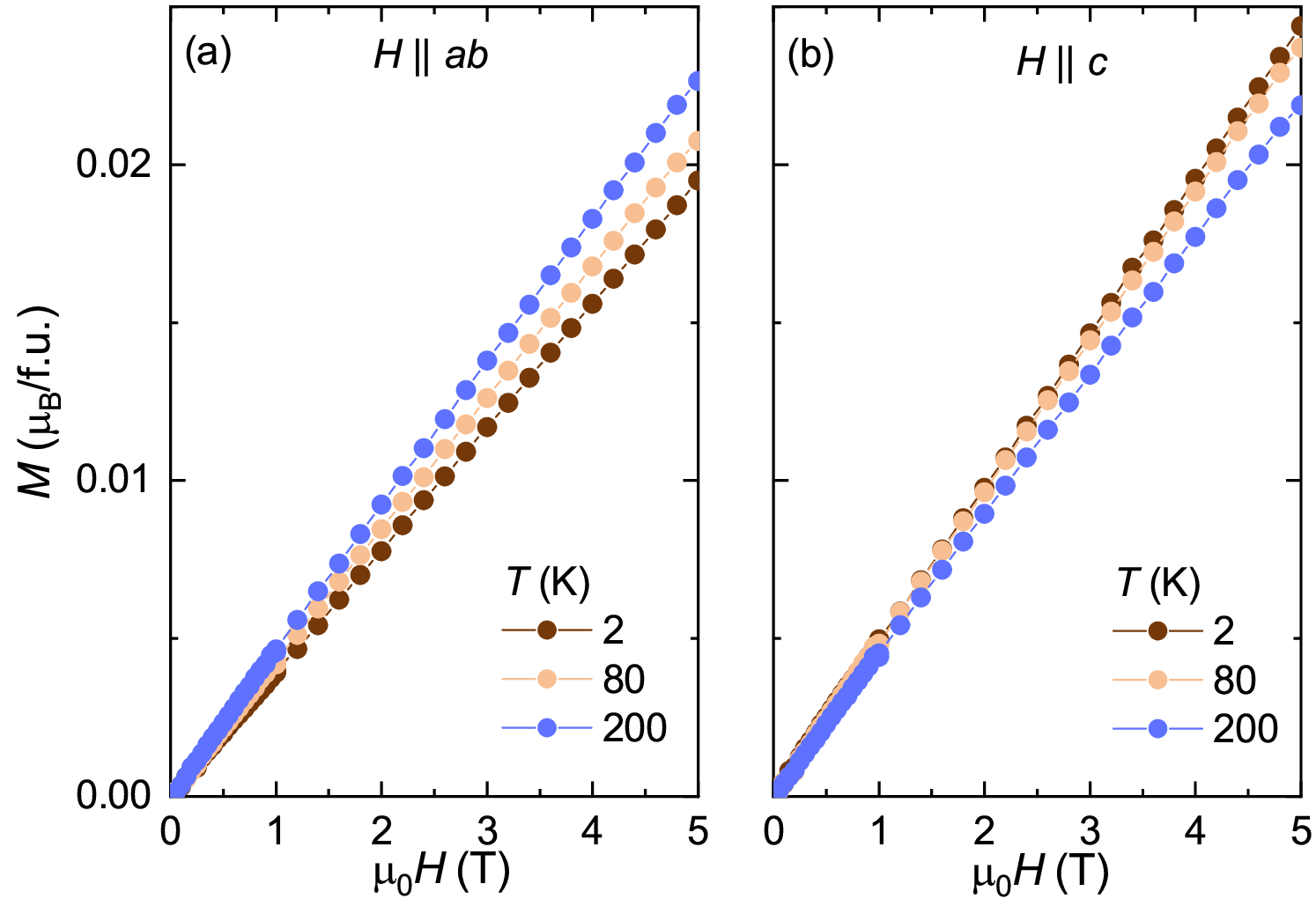}
		\vspace{-12pt} \caption{\label{MH} Magnetization as a function of the magnetic field applied along the (a) $ab$-plane and (b) along the $c$-axis, at various temperatures.}
		\vspace{-12pt}
	\end{figure} 
	
	The field-dependent magnetization $M(H)$ (Fig.~\ref{MH}) exhibits linear behavior for both field orientations in both antiferromagnetic and paramagnetic states. The induced moment remains small, reaching only $\sim$0.02~$\mu_{\rm{B}}$ at 5~T, which is significantly lower than the effective moment of 4~$\mu_{\rm{B}}$ derived from the Curie-Weiss analysis. 
	
\begin{figure}
		\includegraphics[angle=0,width=0.42\textwidth]{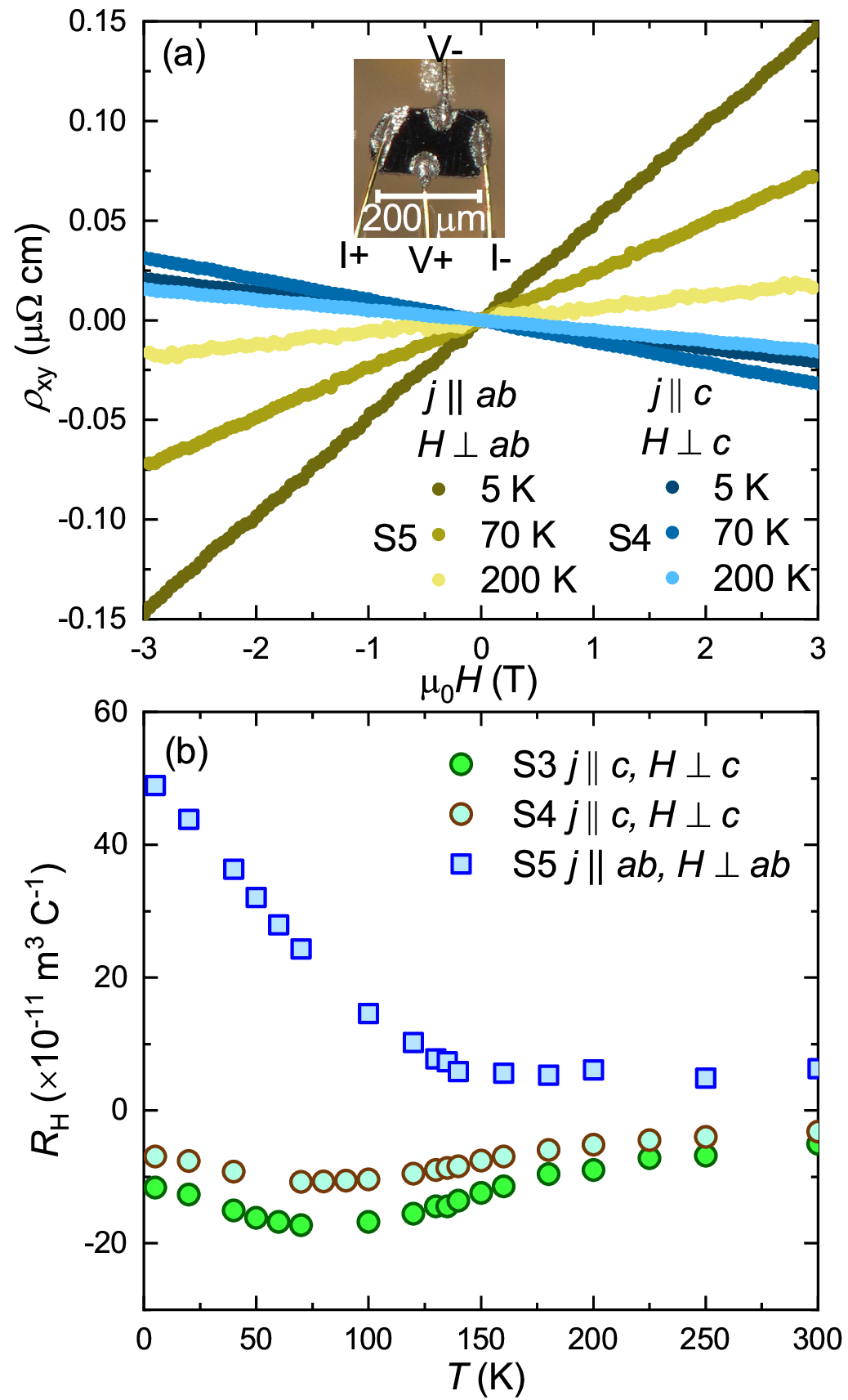}
		\vspace{-12pt} \caption{\label{CRAHall} (a) Hall resistivity of CrRhAs. Data of two samples were shown. The current was applied along the $ab$-plane ($c$-axis) and the magnetic field was applied perpendicular to the $ab$-plane ($c$-axis) for S5 (S4). Inset shows the Hall configuration of sample S5. (b) Temperature evolution of the Hall coefficient for three samples.}
		\vspace{-12pt}
	\end{figure}

	Figure~\ref{CRAHall}(a) displays the Hall resistivity $\rho_{\rm{xy}}$ of CrRhAs for two distinct current and field configurations. Two key observations emerge: (i) $\rho_{xy}$ varies linearly with magnetic field for both orientations, and (ii) the slope of $\rho_{\rm{xy}}$ switches sign depending on the measurement geometry: it is positive for $j \parallel ab$, $H \perp ab$ and negative for $j \parallel c$, $H \perp c$. These characteristics persist across both the antiferromagnetic and paramagnetic states. 
	
	The Hall coefficient $R_{\rm H}$, extracted from linear fits to $\rho_{xy} = R_{\rm H} B$, exhibits a pronounced anisotropic behavior [Fig.~\ref{CRAHall}(b)]. For $j \parallel ab$, $R_{\rm H}$ shows weak temperature dependence above $T_{\rm N} = 150$~K but undergoes a dramatic enhancement (around tenfold) below the magnetic transition, reaching its maximum value at the lowest measured temperature. In contrast, for $j \parallel c$, $|R_{\rm H}|$ increases gradually from 300~K, with an accelerated growth below $T_{\rm N}$ and a broad peak emerging near 75~K. The same behavior is observed in another sample. The striking directional dependence of $R_{\rm H}$ suggests an anisotropic reconstruction of the Fermi surface in the antiferromagnetic state.
	
	\section{IV. Discussion}
\begin{figure}
		\includegraphics[angle=0]{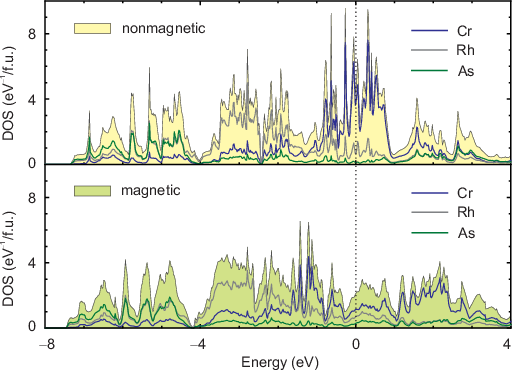}
		\vspace{-12pt} \caption{\label{fig:dos} 
			Density of states for CrRhAs calculated using the PBE functional with (bottom) and without (top) introducing the spin polarization. The Fermi level is at zero energy. }
		\vspace{-12pt}
	\end{figure}

	CrRhAs stands apart from other kagome metals in that its ground state is antiferromagnetic. Antiferromagnetism arises from the mostly localized magnetic moments that can be tentatively assigned to Cr$^{3+}$. It is also instructive to analyze the role of correlation effects in this material, as the Sommerfeld coefficient of $\gamma = 33$\,mJ\,mol$^{-1}$\,K$^{-2}$ appears to be quite high for a $3d$ compound. Indeed, the nonmagnetic DFT calculation (Fig.~\ref{fig:dos}) returns the density of states (DOS) at the Fermi level of $N(E_{\rm{F}})\simeq 4.5$\,eV$^{-1}$/f.u. that corresponds to the broad peak of Cr $3d$ states near the Fermi level. Introducing magnetism splits these states and largely removes them from the Fermi level, resulting in $N(E_{\rm{F}})\simeq 2.1$\,eV$^{-1}$/f.u. for the collinear up-up-down magnetic structure or 2.3\,eV$^{-1}$/f.u. for the nearest-neighbor $120^{\circ}$ magnetic structure. We thus obtain the bare Sommerfeld coefficient of $\gamma_{\rm bare}\simeq 5.2$\,mJ\,mol$^{-1}$\,K$^{-2}$ and the sizable renormalization factor of $m^*/m_e=\gamma/\gamma_{\rm bare}\simeq 6.3$ indicative of moderate electronic correlations in CrRhAs.
	
	Another interesting observation is the large effective moment of $4.3-4.4$\,$\mu_B$ obtained from the Curie-Weiss analysis of the magnetic susceptibility. It resonates with the local magnetic moment of about 3\,$\mu_{\rm{B}}$ on Cr atoms in DFT (PBE). CrRhAs shows a strong tendency toward localization of the Cr $3d$ electrons even without electronic correlations introduced. The itinerant nature of the material should be at least partially attributed to the residual Rh and As states at the Fermi level.
	
	Although magnetic structure of CrRhAs has not been determined experimentally yet, our data suggest a noncollinear spin configuration with spins lying in the $ab$ plane, because the measured $M(T)$ resembles that of noncollinear antiferromagnets~\cite{12JohPRL,19JinPRB,19HudJPCM}. Such a magnetic structure has been confirmed in a sister compound, FeCrAs~\cite{19JinPRB,19HudJPCM}. It was also shown to be energetically favorable in CrRhAs~\cite{23HuaQM}. 
	A natural question at this juncture is why the Hall resistivity of the material is essentially linear, as in ordinary metals, and lacks any additional contributions that would typically occur in metals with complex magnetic order. 
	
	Before discussing the Hall response, we briefly comment on the possible extrinsic effects related to sample imperfections, such as the weak Cr/Rh site mixing. First, the antiferromagnetic transition at $T_{\rm N}$ is consistently and clearly resolved in resistivity, heat capacity, and magnetization measurements, confirming the high quality of our single crystals. Second, the Hall response is well reproducible across different samples. Third, the pronounced evolution of the Hall coefficient emerges only below $T_{\rm N}$, strongly suggesting that it is intrinsically connected to the magnetic ordering rather than arising from extrinsic effects, such as impurities or disorder.
	
	
	One unusual feature of the Hall response in CrRhAs is its sign change upon swapping the directions of current and applied magnetic field. In the simplest form, the Hall coefficient $R_{\rm{H}}$ provides a direct measure of the carrier concentration $n$ through the relation:
	\begin{equation}
		R_{\rm{H}} = \frac{1}{n q}
	\end{equation}
	where $q$ is the elementary charge. This relation usually breaks down in multiband systems. Thus, a two-band version is conventionally adopted,
	\begin{equation}
		R_{\rm{H}} = \frac{p\mu_h^2 - n\mu_e^2}{e(p\mu_h + n\mu_e)^2}
	\end{equation}
	where $n$ ($p$) and $\mu_e$ ($\mu_h$) represent the electron (hole) concentrations and mobilities, respectively. However, under low-field conditions ($\omega_c \tau \ll 1$) where the Drude relaxation time ($\tau$) is significantly shorter than the electron cyclotron period (as in our case), electrons traverse only a small segment of the Fermi surface before undergoing scattering events (e.g., due to phonons or impurities). In this regime, the Hall effect predominantly reveals the anisotropic scattering rate, electron velocity, and effective mass at the specific Fermi surface points, rather than providing a direct information about the true carrier type or the volume of the Fermi surface~\cite{78BanPRB, hurd2012hall, 91OngPRB, 19OuiCI}, especially in highly anisotropic systems. Typically, the Hall response is dominated by the segments of the Fermi surface that have high Fermi velocity and large curvature in semiclassical Boltzmann formalism. Therefore, local Fermi surface geometry is important for interpreting the Hall effect.
	
	\begin{figure}
		\includegraphics[angle=0,width=0.25\textwidth]{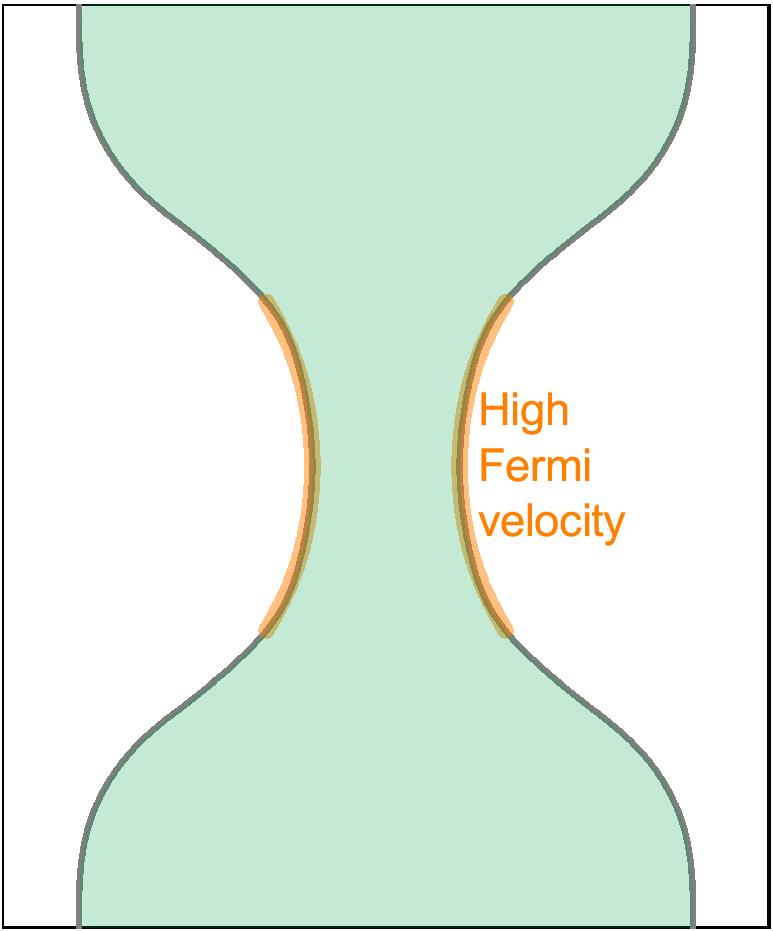}
		\vspace{-12pt} \caption{\label{ConcFS} A schematic Fermi surface with a concave curvature inferred from the DFT results of Ref.~\cite{23HuaQM}. The orange shade indicates the segments with a high Fermi velocity.}
		\vspace{-12pt}
	\end{figure}
	
	In anisotropic systems, the Hall coefficient can display sign reversals originating from Fermi surface topology and competing carrier dynamics~\cite{02EltPRB, 14AraJPSJ, 15UchPRB, 19ZhaPRB, 20MonPRB, 23LuoPRB}. Particularly in systems with multiple carrier types, the dominant contribution to transport properties becomes direction-dependent, with anisotropic carrier mobilities determining the observed Hall response based on the current orientation relative to the crystallographic axes. This sign reversal can even occur in anisotropic single-band systems~\cite{19HeNM}. DFT predicts a multiband nature of CrRhAs, and a highly anisotropic band topology in terms of the Fermi velocity and the curvature of the Fermi surface~\cite{23HuaQM}. An inspection of the electronic bands of CrRhAs reveals a Fermi surface with a concave shape in combination with a high Fermi velocity (as schematically shown in Fig.~\ref{ConcFS}, and Fig. 8h in Ref.~\cite{23HuaQM}). This peculiar topology has been implied as the main driver of the sign change of the Hall coefficient, as observed in NaSn$_2$As$_2$~\cite{19HeNM}, and in LaRh$_6$In$_4$~\cite{23LuoPRB}, which could also qualitatively explain the similar observation in CrRhAs.    
	
	
	
	In the paramagnetic phase of CrRhAs, the Hall coefficient exhibits a relatively weak temperature dependence—a behavior typical of conventional metals, which originated in phonon scattering in Bloch-Boltzmann theory~\cite{91OngPRB}. A pronounced change in the Hall coefficient between the paramagnetic and antiferromagnetic phases has been reported in other antiferromagnetic systems~\cite{69RhyJAP, 09MunPRB, 15FenPRB, 24ShiPRB}. Such a transition is often attributed to either Fermi surface reconstruction (e.g., via the opening of a spin-density-wave gap), or changes in the anisotropic scattering rate (e.g., due to the scattering from magnons or a spin-dependent scattering). In correlated metallic systems, these effects are often difficult to disentangle, and the observed behavior likely originates from a combination of the two contributions. The underlying mechanism in CrRhAs requires further dedicated investigation of the Fermi surface to elucidate the origin of these Hall effect anomalies. 
	
	Finally, we note that CrRhAs does not show any nonlinear contribution to the Hall effect. This observation does not rule out the part of AHE, which is proportional to magnetization, because magnetization of CrRhAs evolves linearly with the field, and the corresponding AHE contribution is indistinguishable from the ordinary Hall effect. On the other hand, our data do not support the presence of an intrinsic AHE that would be caused, for example, by the Berry curvature. The absence of such a contribution should be traced back to the magnetic structure of CrRhAs, which requires further dedicated investigation.  
	
	\section{V. Summary}
	
	In summary, we have grown single crystals of the kagome antiferromagnet CrRhAs and uncovered several features of its low-temperature behavior: i) In-plane spin direction in the antiferromagnetically ordered state, most likely with a non-collinear configuration arising from the frustrated kagome geometry; ii) Absence of nonlinear contributions to the Hall effect; iii) Sign change of the Hall coefficient depending on the directions of the applied field and current; iv) Pronounced change in the Hall coefficient at the onset of antiferromagnetic order. Our results elucidate the strong interplay between magnetism and electronic transport in antiferromagnetic CrRhAs, highlighting its distinctive behavior among metallic kagome magnets and providing an interesting counterpart to the more common ferromagnetic kagome metals.
	
	\section{Acknowledgments}
	We acknowledge fruitful discussions with Igor Mazin. This work was funded by the Deutsche Forschungsgemeinschaft (DFG, German Research Foundation) -- TRR 360 -- 492547816 (subproject B3). B.S. was supported by the Alexander von Humboldt Foundation. We thank ESRF for providing the beamtime for this project and acknowledge technical assistance by Victoria Ginga and Andy Fitch. The authors gratefully acknowledge the use of computing resources of the ALCC HPC cluster (Institute of Physics, University of Augsburg).
	
	\section{Data availability}
	Experimental data associated with this study are available from Ref.~\cite{CRAdata}.

		\paragraph*{Note added.}
		While this manuscript was under review, a related study on polycrystalline CrRhAs was reported~\cite{26ShaPRB}, showing transport and magnetic properties qualitatively similar to those observed in our single crystals.

\end{document}